\documentclass[pre,twoside,twocolumn,hyperref,byrevtex,superscriptaddress]{revtex4}

\usepackage{graphicx,epsfig,verbatim,enumerate}
\usepackage{amssymb}
\usepackage{ifthen}
\newboolean{twocolswitch}

\newcommand{\www}[1]{\url{#1}}

\setboolean{twocolswitch}{true}

\begin{document}

\title{
Identity and search in social networks
}

\markboth{IDENTITY AND SEARCH IN SOCIAL NETWORKS}{WATTS, DODDS, \& NEWMAN}

\author{
  \firstname{Duncan J.}
  \surname{Watts}
  }
\email{djw24@columbia.edu}
\affiliation{
        Department of Sociology,
        Columbia University,
        New York, NY 10027.
        }
\affiliation{
        Columbia Earth Institute, 
        Columbia University,
        New York, NY 10027.
        }
\affiliation{
        Santa Fe Institute, 
        1399 Hyde Park Road, 
        Santa Fe, NM 87501
        }

\author{
  \firstname{Peter Sheridan}
  \surname{Dodds}
  }
\email{p.s.dodds@columbia.edu}
\affiliation{
        Columbia Earth Institute, 
        Columbia University,
        New York, NY 10027.
        }

\author{
  \firstname{M. E. J.}
  \surname{Newman}
  }
\email{mark@santafe.edu}
\affiliation{
        Santa Fe Institute, 
        1399 Hyde Park Road, 
        Santa Fe, NM 87501
        }

\date{\today}

\begin{abstract}
Social networks have the surprising property
of being ``searchable'':
ordinary people are capable of directing messages through 
their network of acquaintances 
to reach a specific but distant target person in only a few steps.
We present a model
that offers an explanation of social network searchability 
in terms of recognizable personal identities defined
along a number of social dimensions.  
Our model defines a class of searchable networks
and a method for searching them 
that may be applicable to many network search problems including 
the location of data files in peer-to-peer networks,
pages on the World Wide Web, and information in
distributed databases.
\end{abstract}


\maketitle

In the late 1960's,
Travers and Milgram~\cite{travers69} conducted an experiment in which randomly
selected individuals in Boston, Massachusetts, and Omaha, Nebraska,
were asked to direct letters to a target person in Boston, each forwarding
his or her letter to a single acquaintance whom they
judged to be closer than themselves to the target.  Subsequent recipients
did the same.  
The average length of the resulting acquaintance chains for the letters
that eventually reached the target (roughly $20\%$) was approximately six.
This reveals not only that 
short paths exist~\cite{watts98,strogatz2001}
between individuals in a large social network but that
ordinary people can find these short paths~\cite{kleinberg2000}.
This is not a trivial statement, since people 
rarely have more than local knowledge about the network.  
People know who their friends are.  
They may also know who some of their friends' friends are.
But no one knows the identities of the entire chain of
individuals between themselves and an arbitrary target.  

The property of being able to find a target quickly, 
which we call searchability, has been shown to
exist in certain specific classes of networks
that either possess a certain fraction of hubs (highly
connected nodes which, once reached, can distribute messages to all parts
of the network~\cite{barabasi99,adamic2001,kim2001u})
or are built upon an underlying geometric lattice
which acts as a proxy for ``social space''~\cite{kleinberg2000}.  
Neither of these network types, however, is
a satisfactory model of society.

In this paper, we present a model for a social network that is based
upon plausible social structures and offers an explanation for the
phenomenon of searchability.  Our model follows naturally from six
contentions about social networks.

\begin{figure}[tbp]
  \begin{center} 
    \epsfig{file=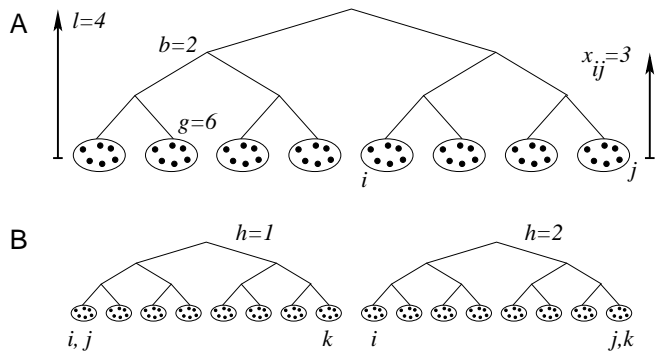,width=0.48\textwidth}
    \caption{
      \textbf{(A)}
      Individuals (dots) belong to groups (ellipses) which in turn belong to
      groups of groups and so on giving rise to a hierarchical
      categorization scheme.  In this example, groups are composed
      of $g=6$ individuals and the hierarchy has $l=4$ levels with
      a branching ratio of $b=2$.  
      Individuals in the same group are considered to be a distance
      $x=1$ apart and the maximum separation of two individuals is $x=l$.  
      The example individuals $i$ and $j$ belong to a category
      two levels above that of their respective groups and the
      distance between them is $x_{ij}=3$.
      Individuals each have $z$ friends in the model
      and are more likely to be 
      connected with each other the closer their groups are.
      \textbf{(B)}
      The complete model has many hierarchies indexed by $h=1\ldots H$, and the
      combined social distance $y_{ij}$ between nodes $i$ and $j$ is taken to be the
      minimum ultrametric distance over all hierarchies $y_{ij} = \min_h x_{ij}^h$.
      The simple example shown
      here for $H=2$ demonstrates that social distance can
      violate the triangle inequality: $y_{ij}=1$ since $i$ and $j$
      belong to the same group under the first hierarchy
      and similarly $y_{jk}=1$ but $i$ and $k$ remain distant
      in both hierarchies giving $y_{ik}=4>y_{ij}+y_{jk}=2$.
      }
    \label{fig.search:hierarchy} 
  \end{center}
\end{figure}

\textbf{1.} 
Individuals in social networks are endowed not only with network ties, but
identities~\cite{white92}: sets of characteristics which they
attribute to themselves and others by virtue of their association with, and
participation in, social groups~\cite{simmel02,nadel57}.  
The term group refers to any collection 
of individuals with which some well-defined
set of social characteristics is associated.

\textbf{2.} 
Individuals break down, or cluster, the world
hierarchically into a series of layers, where the top layer
accounts for the entire world and each successively deeper layer represents
a cognitive division into a greater number of increasingly specific
groups.  In principle, this process of distinction by division can be
pursued all the way down to the level of individuals, at which point each
person is uniquely associated with his or her own group.  For purposes of
identification, however, people do not typically do this, instead
terminating the process at the level where the corresponding group size $g$
becomes cognitively manageable.  
Academic departments, for example, are
sometimes small enough to function as a single group, but tend to
split into specialized sub-groups as they grow larger.  
A reasonable upper bound on group size~\cite{simmel02} is $g \simeq 100$, 
a number which we incorporate into our model 
(Fig.~\ref{fig.search:hierarchy}A).
We define
the similarity $x_{ij}$ between individuals $i$ and $j$ as the height of their
lowest common ancestor level in the resulting hierarchy, setting $x_{ij}=1$
if $i$ and $j$ belong to the same group.
The hierarchy is fully characterized by depth $l$ and constant branching
ratio $b$.  The hierarchy is a purely cognitive
construct for measuring social distance and not an actual network.  The
real network of social connections is constructed as follows.

\textbf{3.} 
Group membership, in addition to defining individual identity, is 
a primary basis for
social interaction~\cite{nadel57,breiger74}, and therefore acquaintanceship. 
As such, the probability of acquaintance between individuals $i$ and $j$
decreases with decreasing similarity of the groups to which
they respectively belong.  We model this by choosing an individual $i$ at
random and a link distance $x$ with probability $p(x)=c\exp\{-\alpha x\}$,
where $\alpha$ is a tunable parameter, and $c$ is a normalizing constant. 
We then choose a second node $j$ uniformly among all nodes that are
distance $x$ from $i$, repeating this process until we have constructed a
network in which individuals have an average number of friends $z$.  The
parameter $\alpha$ is therefore a measure of
homophily---the tendency of like to associate with like.
When $e^{-\alpha} \ll 1$, 
all links will be as short as possible, and individuals will
only connect to those most similar to themselves (i.e., members of their
own bottom-level group), yielding a completely homophilous world of
isolated cliques.  By contrast, when $e^{-\alpha} = b$, any individual
is equally likely to interact with any other, yielding a uniform random
graph~\cite{bollobas85} in which the notion of individual similarity or 
dissimilarity has become irrelevant.

\textbf{4.} 
Individuals hierarchically cluster
the social world in more than one way 
(for example, by geography and by occupation).  
We assume that these categories are
independent, in the sense that proximity 
in one does not imply proximity in another.  
For example, two people may live in the 
same town but not share the same profession. 
In our model, we represent each such
social dimension by an independently partitioned hierarchy.
A node's identity is then defined as an
$H$-dimensional coordinate vector $\vec{v}_i$, where $v_i^h$ is the
position of node $i$ in the $h$th hierarchy, or dimension.  
Each node $i$ is randomly assigned a coordinate 
in each of $H$ dimensions, and is then
allocated neighbors (friends) 
as described above, where now it randomly
chooses a dimension $h$ (e.g. occupation) to use for each tie.  
When $H=1$ and $e^{-\alpha} \ll 1$, the density
of network ties must obey the constraint $z<g$.

\textbf{5.} 
Based on their perceived similarity with other nodes,
individuals construct a measure of ``social 
distance'' $y_{ij}$, which we define as the minimum ultrametric distance 
over all dimensions between two nodes $i$ and $j$; i.e.,
$y_{ij}=\min_h x_{ij}^{h}$.  This minimum metric captures the
intuitive notion that closeness in only a single dimension is sufficient to
connote affiliation (for example, geographically and ethnically distant
researchers who collaborate on the same project).  A consequence of this
minimal metric, depicted in Fig.~\ref{fig.search:hierarchy}B, is that
social distance violates the triangle inequality---hence it is not a true
metric distance---because individuals $i$ and $j$ can be close in dimension
$h_1$, and individuals $j$ and $k$ can be close in dimension $h_2$, yet $i$
and $k$ can be far apart in both dimensions.

\textbf{6.}
Individuals forward a message to a single neighbor given only
local information about the network.
Here, we suppose that each node $i$ knows only its
own coordinate vector $\vec{v}_i$, the coordinate vectors $\vec{v}_j$
of its immediate network neighbors, and
the coordinate vector of a given target individual
$\vec{v}_t$, but is otherwise ignorant of the identities or network ties of
nodes beyond its immediate circle of acquaintances.  

Individuals therefore have two kinds of partial information: social distance,
which can be measured globally but which is not a true distance 
and hence can yield misleading estimates; 
and network paths, which generate true
distances but which are known only locally.  
Although neither kind of information alone is sufficient to perform efficient
searches, here we show that a simple algorithm that 
combines knowledge of
network ties and social identity can succeed in directing messages with
efficiency. 
The algorithm we implement is the same greedy algorithm Milgram
suggested: each member $i$ of a message chain forwards the message to its
neighbor $j$ who is perceived to be closer to the target $t$
in terms of social distance; 
that is, $y_{jt}$ is minimized over 
all $j$ in $i$'s network neighborhood.

\begin{figure}[tbp]
  \begin{center}
    \epsfig{file=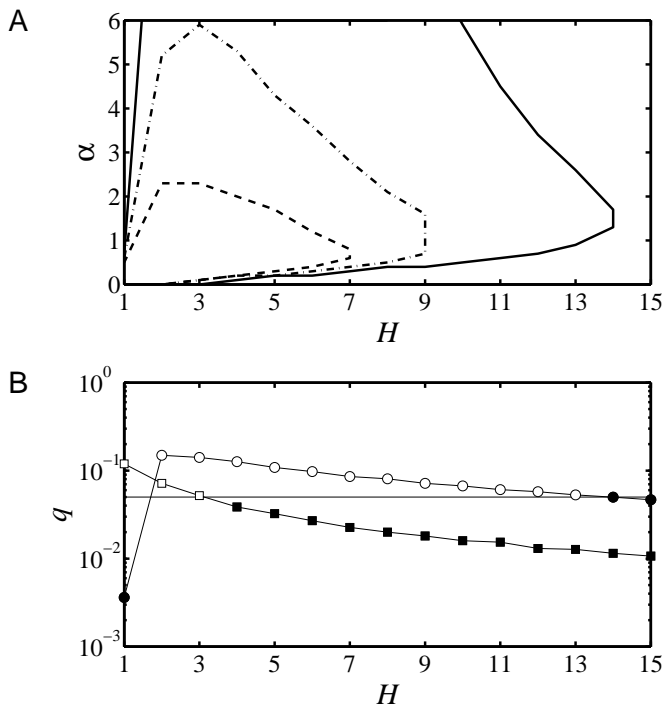,width=0.49\textwidth}
    \caption{ 
      \textbf{(A)}
      Regions in $H$-$\alpha$ space where
      searchable networks exist for varying 
      numbers of individual nodes $N$
      (probability of message failure $p=0.25$,
      branching ratio $b=2$, group size $g=100$, average degree $z=g-1=99$,
      $10^5$ chains sampled per network).
      The searchability criterion is that the
      probability of message completion $q$ must be at least $r=0.05$.
      The lines correspond to
      boundaries of the searchable network region for $N=102400$ (solid),
      $N=204800$ (dot-dash), and $N=409600$ (dash).
      The region of searchable networks shrinks
      with $N$, vanishing at a finite value of $N$ which
      depends on the model parameters.
      Note that $z<g$ is required to explore $H$-$\alpha$ space
      since for $H=1$ and $\alpha$ sufficiently large, 
      an individual's neighbors must all be contained 
      within their sole local group.
      \textbf{(B)}
      Probability of message completion $q(H)$ when
      $\alpha=0$ (squares) and $\alpha=2$ (circles)
      for the $N=102400$ data set used in \textbf{a}.
      The horizontal line shows the position of the threshold $r=0.05$.
      Open symbols indicate the network is searchable ($q \geq r$)
      and closed symbols mean otherwise.
      For $\alpha=0$, searchability degrades
      with each additional hierarchy.  
      For the homophilous case of $\alpha=2$ with a single hierarchy,
      less than one percent of all searches find their target 
      ($q \simeq 0.004$).
      Adding just one other
      hierarchy increases the success rate to $q \simeq 0.144$
      and $q$ slowly decreases with $H$ thereafter.
      }
    \label{fig.search:basicresults}
  \end{center}
\end{figure}

Our principal objective is to determine the conditions
under which the average length $\langle L \rangle$ of a 
message chain connecting a randomly selected
sender $s$ to a random target $t$ is small.
Although the term small has recently
been taken to mean that $\langle L \rangle$ grows slowly
with the population size $N$~\cite{newman99b,kleinberg2000b},
Travers and Milgram found only
that chain lengths were short.
Furthermore, these message chains 
had to be short in an absolute sense
because at each step,
they were observed to terminate
with probability $p \simeq 0.25$~\cite{travers69,white70}.
We therefore adopt a more realistic, 
functional notion of efficient search, defining 
for a given message failure probability $p$,
a \emph{searchable network} 
as any network for which $q$, the probability
of an arbitrary message chain reaching its target, is
at least a fixed value $r$.  In terms of chain length,
we formally require $q = \langle(1-p)^{L}\rangle \geq r$,
and from this we can obtain an estimate of the maximum required $\langle L \rangle$ using
the approximated inequality $\langle L \rangle \leq \ln{r}/\ln{(1-p)}$.
For the purposes of this paper,
we set $r=0.05$ and $p=0.25$ giving
the stringent requirement that
$\langle L \rangle \leq 10.4$ independent of the population size $N$.
Fig.~\ref{fig.search:basicresults}A
presents a typical phase diagram in $H$ 
and $\alpha$ outlining the searchable network region 
for several choices of $N$, $g=100$, and $z=g-1=99$.

Our main result is that searchable networks occupy a broad region of
parameter space $\left(\alpha,H\right)$ which, as we argue below,
corresponds to choices of the model parameters that are the most
sociologically plausible.  Hence our model suggests that 
searchability is a generic property of real-world social networks. We
support this claim with some further observations, and demonstrate
that our model can account for Milgram's experimental findings.

First, we observe that almost all searchable networks display
$\alpha>0$ and $H>1$, consistent with the notion that individuals are
essentially homophilous (that is, they associate preferentially with
like individuals), but judge 
similarity along more than one social dimension.
Neither the precise degree to which they
are homophilous, nor the exact number of dimensions they choose to
use, appear to be important---almost any reasonable choice will do.
The best performance, over the largest interval
of $\alpha$, is achieved for $H=2$ or $3$---an interesting
result in light of empirical 
evidence~\cite{bernard88} that individuals
across different cultures in small-world experiments typically
utilize two or three dimensions when forwarding a message.

Second, as Fig.~\ref{fig.search:basicresults}B shows, while increasing
the number of independent dimensions from $H=1$ yields a dramatic
reduction in delivery time for values of $\alpha > 0$, this
improvement is gradually lost as $H$ is increased further.  Hence the
window of searchable networks in Fig.~\ref{fig.search:basicresults}A
exhibits an upper boundary in $H$.  Because ties
associated with any one dimension are allocated independently with
respect to ties in any other dimension, and because 
for fixed average degree $z$, larger $H$
necessarily implies fewer ties per dimension, the network ties become
less correlated as $H$ increases.  In the limit of large $H$, the
network becomes essentially a random graph (regardless of $\alpha$)
and the search algorithm becomes a random walk.  Effective
decentralized search therefore requires a balance (albeit a highly
forgiving one) of categorical flexibility and constraint.

\begin{figure}[tbp]
  \begin{center}
    \epsfig{file=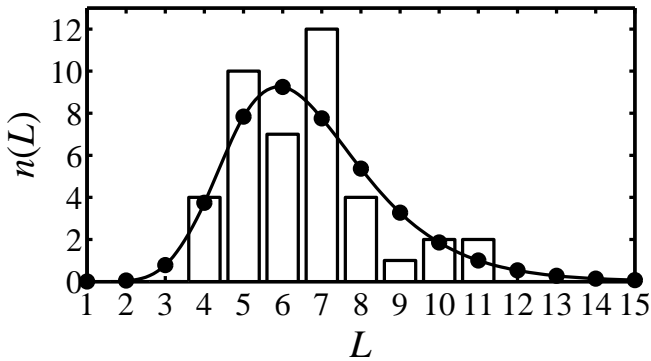,width=0.49\textwidth}
    \caption{ 
      Comparison between $n(L)$, the number of completed chains of 
      length $L$, taken from the original
      small-world experiment~\cite{travers69} (bar graph)
      and from an example of our model with $N=10^8$ individuals 
      (filled circles with the line being a guide for the eye).
      The experimental data shown are for the 42 completed chains
      that originated in Nebraska.  (We have excluded the
      24 completed chains that originated in Boston
      as this would correspond to $N \simeq 10^6$.)
      The model parameters are $H=2$, $\alpha=1$, $b=10$,
      $g=100$, and $z=300$; message attrition rate is set at 25\%;
      $n(L)$ for the model is compiled from $10^6$ random chains
      and is normalized to match 
      the 42 completed chains that started in Nebraska.
      The average chain length of Milgram's experiment
      is approximately 6.5 while 
      the model yields $\langle L \rangle \simeq 6.7$.
      The distributions compare well:
      a two-sided Kolmogorov-Smirnov test yields 
      a p-value $P \simeq 0.57$ 
      while for a $\chi^2$ test, $\chi^2 \simeq 5.46$
      and $P \simeq 0.49$ (seven bins).
      (A large value of $P$ supports the 
      hypothesis that the distributions are similar.)
      Even without attrition, the model's average
      search time is $\langle L \rangle \simeq 8.5$ and
      the median chain length is 8.  The model does not entirely match
      the experimental data since the former requires approximately 360
      initial chains to achieve 42 completions as compared to 196.
      } 
    \label{fig.search:milgram} 
  \end{center}
\end{figure}

Finally, by introducing parameter choices that are consistent with Milgram's 
experiment ($N=10^8$, $p=0.25$) \cite{travers69}, as well as with subsequent 
empirical findings ($z=300$, $H=2$)\cite{killworth78,bernard88}, we can 
compare the distribution of chain lengths in our model with those of 
Travers and Milgram \cite{travers69} for plausible values of $\alpha$ and $b$. 
As Fig.~\ref{fig.search:milgram} shows,
we obtain $\langle L \rangle \simeq 6.7$ for $\alpha=1$ 
and $b=10$, indicating that our
model captures the essence of the real small-world problem.

Although sociological in origin, our model
is relevant to a broad class of decentralized search problems, such as
peer-to-peer networking, in which centralized servers are excluded either by
design or by necessity, and where broadcast-type searches (i.e., forwarding 
messages to all neighbors rather than just one) are ruled out due to 
congestion constraints \cite{adamic2001}.  
In essence, our model applies to any data
structure in which data elements exhibit quantifiable characteristics
analogous to our notion of identity, and similarity between two
elements---whether people, music files, web pages, or research
reports---can be judged along more than one dimension.
One of the principal difficulties with designing robust
databases~\cite{manneville99} is the absence of a unique classification
scheme which all users of the database can apply consistently to place and
locate files.  Two musical songs, for example, can be similar because they
belong to the same genre or because they were created in the same year. 
Our model transforms this difficulty into an asset,
allowing all such classification schemes to exist simultaneously, and
connecting data elements preferentially to similar elements in multiple
dimensions.  Efficient decentralized searches can then be conducted
utilizing simple, greedy algorithms providing only that the characteristics
of the target element and the current element's immediate neighbors are
known.

\begin{acknowledgments}
The authors thank Jon Kleinberg for beneficial discussions.
This work was funded in part by the National
Science Foundation under grant numbers SES-00-94162
and DMS--0109086, the Intel Corporation, and
the Columbia University Office of Strategic Initiatives.
\end{acknowledgments}

\appendix

\end{document}